\documentclass[twocolumn,showpacs,prb]{revtex4}
\usepackage{bm}
\usepackage{epsfig}

\newcommand{\nix}[1]{}

\begin{document}

\title{
Pure spin currents induced by spin-dependent scattering processes
in SiGe quantum well structures}
\author{S.D.~Ganichev$^{1}$, S.N.~Danilov$^1$, V.V.~Bel'kov$^{1,2}$,
S.~Giglberger$^1$, S.A.~Tarasenko$^2$, E.L.~Ivchenko$^2$,
D.~Weiss$^1$, W.~Jantsch$^3$, F.~Sch\"{a}ffler$^3$, D.~Gruber$^3$,
and W.~Prettl$^1$}
\affiliation{$^1$Fakult\"{a}t Physik, Universit\"{a}t Regensburg,
93040 Regensburg, Germany}
\affiliation{$^2$A.F.~Ioffe Physico-Technical Institute, Russian
Academy of Sciences, 194021 St.~Petersburg, Russia}
\affiliation{$^3$Institut f{\"u}r Halbleiter und
Festk\"{o}rperphysik, Johannes-Kepler Universit\"{a}t Linz, A-4040
Linz, Austria}


\begin{abstract}
We show that spin-dependent electron-phonon interaction in the
energy relaxation of a two-dimensional electron gas results in equal and
oppositely directed currents in the spin-up and spin-down
subbands yielding a pure spin current. In our experiments on SiGe
heterostructures the pure spin
current is converted into an electric current applying a magnetic
field that lifts the cancellation of the two partial charge flows. A
microscopic theory of this effect, taking account of the asymmetry
of the relaxation process, is developed being in a good agreement
with the experimental data.
\end{abstract}

\pacs{73.21.Fg, 72.25.Fe, 78.67.De, 73.63.Hs}

\maketitle

\newpage

Lately, there is much interest in the use of the spin of carriers
in semiconductor quantum well (QW) structures together with their
charge to realize novel concepts like spintronics and
spin-optoelectronics.~\cite{Zutic04review} The transport of the
spin of charge carriers in semiconductor nanostructures is one of
the key problems in this field. Among the necessary conditions to
realize spintronics devices there are a high spin polarization in QWs
and a large spin splitting of subbands. The latter is important to
control spins by an external electric field via the Rashba
effect.~\cite{Bychkov84p78} While most of the investigations aimed
at spintronics and spin-optoelectronics have been carried out on
III-V compounds, some  recent results obtained on non-magnetic
SiGe nanostructures applying electron spin resonance
(ESR)~\cite{Wilamowski2002p195315,Tyryshkin2003} and the circular
photogalvanic effect (CPGE),~\cite{Ganichev2002b,Belkov2003} 
demonstrated that this material may be a promising system for
spin-based electronics. ESR and CPGE data show that spin
relaxation times in SiGe QWs can be sufficiently
long,~\cite{Wilamowski2002p195315,Tyryshkin2003,Tahan2002,Fanciulli2003,Wilamowski2004p35328,Glazov2004,Tahan2005} 
that the spin degeneracy is
lifted,~\cite{Wilamowski2002p195315,Ganichev2002b,Belkov2003,Tahan2005,Sherman2003} 
that the $g$-factor is tunable by  crystallographic direction,
electron density, Ge-content, kinetic energy of free carriers and
electric
current~\cite{Wilamowski2002p195315,Jantsch2002p504,Kiselev2003,Malissa2004p1739,Truitt2004,Wilamowski2006}
and that  spin manipulation can be achieved by means of the
spin-echo method.~\cite{Tyryshkin2003} 

Here we report on an electrically measured observation of pure
spin currents causing spatial spin separation in SiGe quantum well
structures, allowing  manipulation of spins in this material which
is attractive for high-speed electronics and spintronics. Spin
currents recently attracted rapidly growing interest since they
can provide new tools for the realization of all-electric
non-magnetic semiconductor spintronics. Various phenomena
comprising charge photocurrents driven by the spin degree of
freedom~\cite{PRL01,Nature02,Ganichev03p935,Belkov283p2003,Ivchenkobook2,Bieler05,GanichevPrettl,Yang06,Huebner2003}
and spin separation caused by pure spin
currents~\cite{Stevens2003,Sipe,TI_jetplett,Awschalom,Wunderlich,Ganichev06zerobias}
were reported. Most of these phenomena originate from the well
known lifting  of spin degeneracy. The latter causes the band
structure to split into spin-up and spin-down branches described
by  linear in wave vector $\bm{k}$ terms in the Hamiltonian due to
structure inversion asymmetry (SIA) or bulk inversion asymmetry
(BIA). The pure spin currents reported here are caused by less
known spin-dependent  electron scattering
processes~\cite{bulli,ac_field} which generate  a pure spin
current, causing spin separation in a similar way as in the spin
Hall
effect.~\cite{Awschalom,Wunderlich,Dyakonov71,Hirsch99,Tarasenko} 
In contrast to the latter, no external bias needs to be
applied.~\cite{Ganichev06zerobias} 

Spin separation due to spin-dependent scattering in gyrotropic
media can be achieved in various ways but all of them must drive
the electron gas into a  nonequilibrium state. One straightforward
method used here is to heat the electron system by terahertz (THz)
or microwave radiation.
\begin{figure*}
\centerline{\epsfxsize 140mm \epsfbox{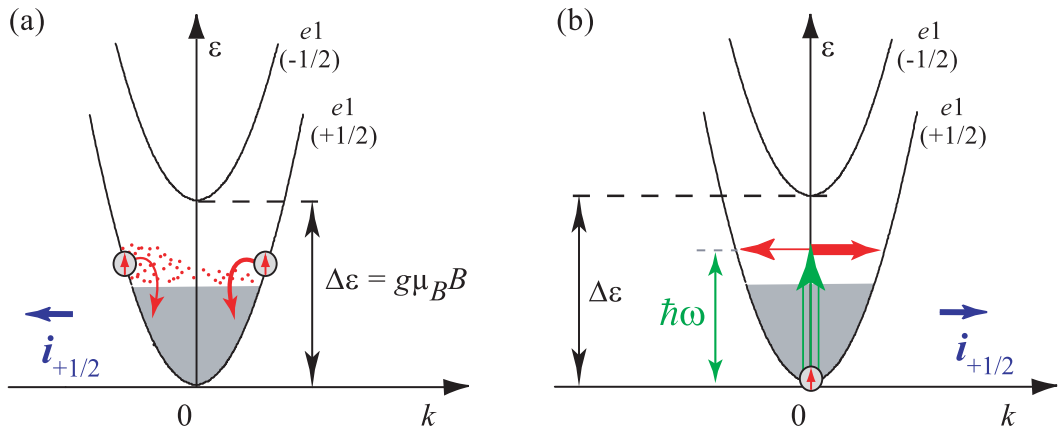}}
\caption{ Microscopic origin of a zero-bias spin separation and
the corresponding magnetic field-induced photocurrent. Zero-bias
spin separation is due to scattering matrix elements linear in
$\bm{k}$ and $\bm{\sigma}$ causing asymmetric  scattering and it
results in spin flows. This process is sketched for the spin-up
subband only and for (a) energy relaxation and (b) excitation via
indirect transitions (Drude-like absorption). Here, scattering is
assumed to have a larger probability for positive $k_x$  than that
for negative $k_x$ as indicated by arrows of different thickness.
Therefore in (a) the energy relaxation rates for positive $k_x$
are larger than  for negative $k_x$ and in (b) the rates of
optical transitions for opposite wave vectors are different. This
imbalance leads to a net spin-up electron flow. In the spin-down
subband the picture is mirror symmetric, resulting in  a net
spin-down electron flow of opposite direction. Thus at zero
magnetic field a spin current is generated. The corresponding
electric currents have equal magnitudes and therefore  cancel each
other. An in-plane magnetic field, however, lifts the compensation
of the oppositely directed electron flows yielding a charge
current. } \label{figure1}
\end{figure*}

Figure~\ref{figure1}(a) sketches the
process of energy relaxation of hot electrons for the spin-up
subband ($s=+1/2$)  in a quantum well containing a two-dimensional
electron gas (2DEG). Energy relaxation processes are shown by
curved arrows. Usually, energy relaxation via scattering of
electrons is considered to be spin-independent. In gyrotropic
media, like low-dimensional GaAs structures or asymmetric SiGe QWs
investigated here, however, spin-orbit interaction adds an
asymmetric spin-dependent term to the scattering
probability.~\cite{Ganichev06zerobias} This term in the scattering
matrix element is proportional to components of
$[\bm{\sigma}\times(\bm{k}+\bm{k}^\prime)]$, where $\bm{\sigma}$
is the vector composed of the Pauli matrices, $\bm{k}$ and
$\bm{k}^\prime$ are the initial and scattered electron wave
vectors.~\cite{footnote1}
Due to spin-dependent scattering, transitions to positive and
negative $k_x^\prime$-states occur with different probabilities.
Therefore hot electrons with opposite $k_x$ have different
relaxation rates in the two spin subbands. In
Fig.~\ref{figure1}(a)  this difference is indicated
by arrows of different
thicknesses.
This asymmetry causes an imbalance in the distribution of carriers
 in both subbands ($s=\pm 1/2$) between positive and negative
$k_x$-states.
This in turn yields a net electron flows, $\bm{i}_{\pm 1/2}$,
within each spin subband. Since the asymmetric part of the
scattering amplitude depends on spin orientation, the
probabilities for scattering to positive or negative
$k^\prime_x$-states are inverted for spin-down and spin-up
subbands.
Thus, the charge currents, $\bm{j}_+ = e\bm{i}_{+1/2}$ and
$\bm{j}_- = e\bm{i}_{-1/2}$, where $e$ is the electron charge,
have opposite directions because $\bm{i}_{+1/2} = -\bm{i}_{-1/2}$
and therefore they cancel each other. Nevertheless, a finite spin
current $\bm{J}_{\rm spin} = \frac{1}{2}(\bm{i}_{+1/2} -
\bm{i}_{-1/2})$ is generated since electrons with spin-up and
spin-down move in opposite directions.~\cite{Ganichev06zerobias} 
This leads to a spatial spin separation and spin accumulation at
the edges of the sample.

Similarly, optical excitation of
free carriers by Drude absorption, also involving
electron scattering, is asymmetric and yields spin separation as
sketched in Fig.~\ref{figure1}(b). We described this mechanism in
detail in previous publications.~\cite{Ganichev06zerobias,TI_jetplett} The model
and the theoretical considerations described there can be directly applied to SiGe
structures. Spin
separation due to hot electron energy relaxation, in contrast, was
only briefly addressed in Ref.~32
and is therefore
discussed in greater details here.

As shown in Ref. 32
a pure spin current and
zero-bias spin separation can be converted into a measurable
electric current by application of a magnetic field. Indeed, in a
Zeeman spin-polarized system, the two fluxes $\bm{i}_{\pm 1/2}$,
whose magnitudes depend on the free carrier densities in spin-up
and spin-down subbands, $n_{\pm 1/2}$, respectively, do no longer
compensate each other and hence yield a net electric current (see
Fig.~\ref{figure1}). For the case, where the fluxes $\bm{i}_{\pm
1/2}$ are proportional to the carrier densities $n_{\pm
1/2}$,~\cite{footnote} the charge current is given by
\begin{equation}
\label{current1}
\bm{j} = e (\bm{i}_{+1/2}  + \bm{i}_{-1/2}) = 4 e S \bm{J}_{\rm
spin} \:,
\end{equation}
where $S = \frac{1}{2}(n_{+1/2} - n_{-1/2})/(n_{+1/2} + n_{-1/2})$
is the magnitude of the average spin. An external magnetic field
$\bm B$ results in different populations of the two spin subbands
due to the Zeeman effect. In equilibrium the average spin is given
by
\begin{equation}
\label{spin}
{\bm S} = - \frac{g \mu_B \bm{B}}{4 \bar{\varepsilon}}\:.
\end{equation}
Here $g$ is the electron effective $g$-factor, $\mu_B$  the Bohr
magneton, $\bar{\varepsilon}$  the characteristic electron energy
being equal to the Fermi energy $\varepsilon_F$, or to the thermal
energy $k_B T$, for a degenerate or a non-degenerate 2DEG,
respectively.~\cite{purespincurrent}

To demonstrate the existence of the spin-polarized current
described above we chose the following experimental conditions:
electron gas heating  is achieved by absorption of linearly
polarized THz radiation at normal incidence on a (001)-grown QW.
Spin polarization is achieved by an in-plane magnetic field and
the current is measured both in the directions normal and parallel
to the magnetic field.

The chosen experimental geometry excludes  other effects that are
known to cause photocurrents in (001)-oriented  QWs: since
linearly polarized radiation is used, all helicity-dependent spin
photocurrents, such as the circular photogalvanic
effect~\cite{PRL01} and the spin-galvanic effect,~\cite{Nature02} 
are absent. In addition, photon drag and the linear photogalvanic
effect are forbidden by symmetry for normal incidence on
(001)-grown  heterostructure (see,
\textit{e.g.}, Refs. 20, 22, 24).

The measurements are carried out on $n$-type   SiGe QW structures,
MBE-grown on (001)-oriented Si-substrates. The samples contain a
single, 15~nm wide, strained Si quantum well deposited on a
3~$\mu$m thick graded buffer ramping  to a composition of
Si$_{0.75}$Ge$_{0.25}$. On top of the tensilely strained Si
quantum well a second Si$_{0.75}$Ge$_{0.25}$ barrier is grown
containing 10$^{18}$~cm$^{-3}$ of Sb for  mo\-du\-lation doping. Due
to the one-sided doping, the two-dimensional electron gas channel
has structure inversion asymmetry.

\begin{figure}
\centerline{\epsfysize 65mm \epsfbox{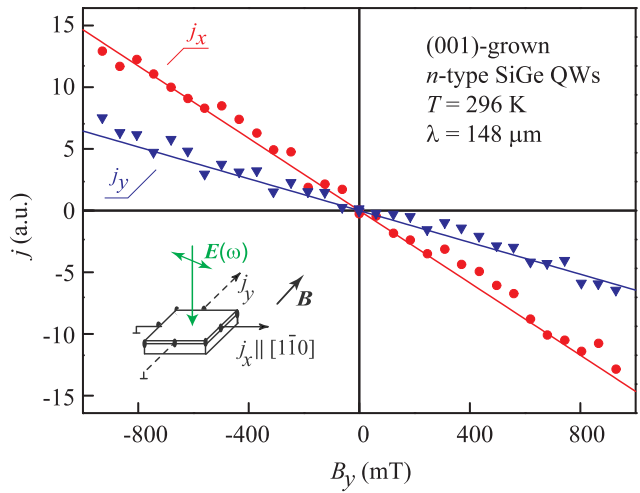}}
\caption{Magnetic field dependence of the photocurrent $j$ measured in
 sample~1 at room temperature with the magnetic field $\bm B$
parallel to the $y$-direction. Radiation of power $P \approx
17$~kW  is applied at normal incidence. Circles show results
obtained for $\bm j \, \bot \, \bm B$, obtained for the radiation
polarized perpendicularly to the magnetic field ($\alpha =
0^\circ$). Triangles show current $\bm j
\parallel \bm B$. These data are given for $\alpha
= 135^\circ$, inset shows the experimental geometry. }
 \label{figure2}
\end{figure}

Two samples with free carrier densities of $2.8 \times
10^{11}$~cm$^{-2}$ (sample~1) and $3.5 \times 10^{11}$~cm$^{-2}$
(sample~2) and corresponding low temperature  mobilities (1.5~K)
of $1.7\times 10^{5}$~cm$^{2}$/Vs  and  $1.0\times
10^{5}$~cm$^{2}$/Vs, respectively, are studied. Two pairs of Ohmic
contacts in the center of the sample edges oriented along $x
\parallel$ [1$\bar{1}$0] and $y \parallel$ [110] have been
prepared (see inset in Fig.~\ref{figure2}).

A high power THz molecular laser, optically pumped by a
TEA-CO$_2$ laser,~\cite{GanichevPrettl} has been used to deliver
100~ns pulses of linearly polarized radiation with a  power of
about 15~kW at a wavelength of $\lambda =$148~$\mu$m. The
radiation causes indirect optical transitions within the lowest
size-quantized subband. The samples are irradiated along the
growth direction. An external magnetic field $B$  up to $1$~T is
applied parallel to the interface plane. The current $\bm j$,
generated by the light in the unbiased devices, is measured via
the voltage drop across a 50~$\Omega$ load resistor in a closed-circuit
configuration. The voltage is recorded with a storage
oscilloscope. Measurements are carried out in a wide temperature
range from liquid helium to room temperature. The measured current
pulses  follow the temporal structure of the applied laser pulses.
In experiment the angle $\alpha$ between the polarization plane of
the light and $x$-axis is varied. This is achieved by a
$\lambda/2$ plate which enables us to vary the angle $\alpha$
between $0^\circ$ and $180^\circ$.

Irradiation of the samples at zero magnetic field does not lead to
any signal as expected from the microscopic mechanism described
above as well as from the phenomenological analysis. A current
response, however, is obtained  when a  magnetic field is
applied. Figure~\ref{figure2} shows the magnetic field dependence
of the photocurrent for two directions: along and perpendicular
to the in-plane magnetic field $\bm B$ (the latter is aligned along $y$). The
current increases linearly with $\bm B$   and changes sign upon
the reversal of the magnetic field direction. This  is in
agreement with the above model because the strength and direction
of the magnetic field affects  the average spin and therefore the
electric current, as given in  Eqs.~(\ref{current1})
and~(\ref{spin}).

\begin{figure}
\centerline{\epsfysize 100mm \epsfbox{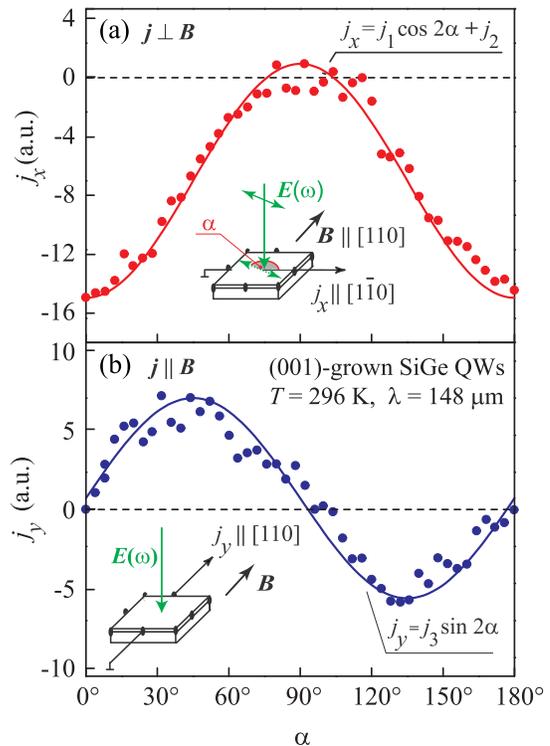}}
%
\caption{Photocurrent in  sample~1  as a function of  $\alpha$.
The sample is excited by normally incident linearly polarized
radiation of power $P \approx 17$~kW. Data are obtained at room
temperature for a magnetic field of $B_y=1$~T. (a) Photocurrent for
$\bm j \, \bot \, \bm B \,\|\,y$. Line: fit of $j_x= j_1 \cos
2\alpha + j_2$. (b) Photocurrent for $\bm j \, \| \, \bm B
\,\|\,y$. Line is fitted to $j_y = j_3 \sin 2\alpha$. The insets show
the experimental geometries. }
 \label{figure3}
\end{figure}

\begin{figure}
\centerline{\epsfxsize 80mm \epsfbox{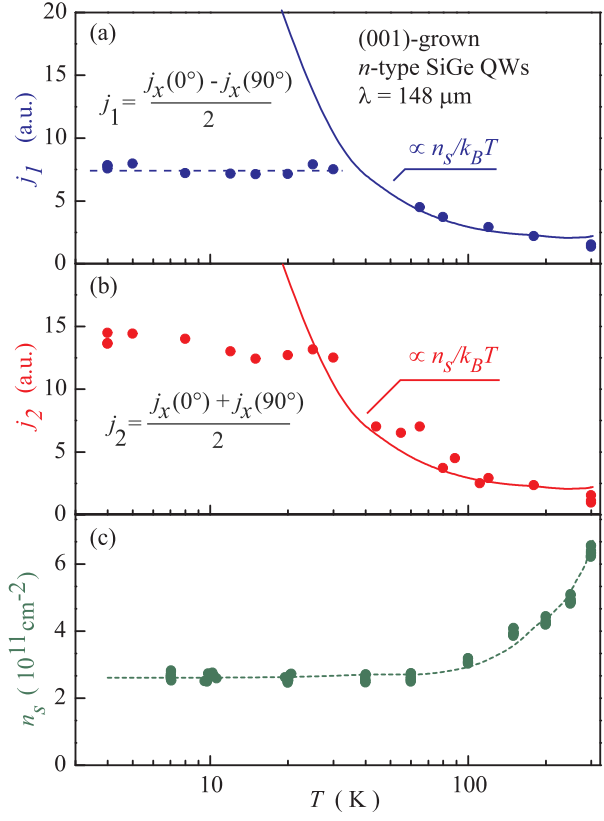}}
\caption{Temperature dependencies  of the transverse photocurrent
and the carrier density $n_s$. Data are obtained for  $B_y=-0.6$~T
applied to sample~1 and an excitation power of $P \approx 5$~kW.
Photocurrents $j_1(T)$ (a) and $j_2(T)$ (b), obtained by
subtracting and adding the currents for the two polarizations:
$\alpha=0^\circ$ and $\alpha=90^\circ$. (c) Temperature dependence
of the carrier density $n_s$. Full lines are fits to $A \cdot
n_s/k_BT$ with a single fitting parameter $A$. The dotted line is meant as a guide for eye.}
 \label{figure4}
\end{figure}

Figures~\ref{figure3} and \ref{figure4} show the dependencies of
the current on polarization and temperature, respectively. We
found that the  polarization dependence of the current  can be
fitted by $j_x = j_1 \cos 2 \alpha + j_2$ for the transverse
geometry and by $j_y = j_3 \sin 2 \alpha $  for the longitudinal
geometry in the whole range of temperature. These polarization
dependencies are  in accordance with the phenomenological theory
of magnetic field induced photocurrents.~\cite{MPGE_jpcm} As shown
in Ref. 32,
the currents $j_1$ and $j_3$ are
caused by indirect optical transitions (Drude absorption) and
therefore exhibit polarization dependence. In contrast, the
current  $j_2$ is driven by energy relaxation of hot electrons and
is therefore independent of the light polarization. It is expected
and observed only for transverse geometry.~\cite{MPGE_jpcm} Using
two fixed polarization directions in the transverse geometry,
$\alpha = 0^\circ$ and $\alpha = 90^\circ$, allows us to extract
both contributions. Adding and subtracting the currents of both
orientations the coefficients $j_1$ (polarization-dependent
amplitude) and $j_2$ (polarization-independent background) can be
obtained by
\begin{equation}
j_1=
\frac{j_x(0^\circ)-j_x(90^\circ)}{2}\,,\,\,\,\,\,\,\,\,\,\,j_2=
\frac{j_x(0^\circ)+j_x(90^\circ)}{2}\,\,.
\end{equation}

Figure~\ref{figure4} shows the temperature dependencies of $j_1$,
$j_2$ and the electron density $n_s$.
Below 100~K both current contributions are almost independent of
temperature, but at temperatures above 150~K the current strength
decreases with increasing temperature.

The peculiar temperature- and polarization- dependencies are a
clear-cut proof that the observed charge current is a result of
imbalanced spin currents. Let us first consider the temperature
behaviour of the polarization-dependent $j_1$ contribution caused
by asymmetric excitation. As has been shown
in Ref 32,
for fixed polarization and certain
scattering mechanism, \textit{e.g.}, phonon or impurity scattering, the
temperature dependence of magneto-photocurrent due to this
mechanism is described by $j_1 / I \propto \tau_p \,\eta (\omega )
S$. Here $I$ stands for the radiation intensity.~\cite{footnote}
Since for Drude absorption and $\omega \tau_p \gg 1$ the
absorbance is given by $\eta (\omega ) \propto n_s /\tau_p $
(see Ref. 42),
the momentum relaxation time, $\tau _p$,
cancels and the temperature dependence of the current reduces to
$j_1 \propto  n_s S$.
At low temperatures $S\propto 1/\varepsilon
_F \propto 1/n_s $ (see Eq.~(\ref{spin}))  and the current  $j_1
\propto  n_s S$ becomes independent of temperature as observed in
our experiment (see dashed line in Fig.~\ref{figure4}(a)).
In additional experiments we change the carrier density at 4.2~K
by visible and near infrared light. By that  the carrier density
(mobility) increases  from $2.8\times 10^{11}$~cm$^{-2}$
($1.7\times 10^5$~cm$^2$/Vs) to $ 3.6 \times 10^{11}$~cm$^{-2}$
($2.2 \times 10^5$~cm$^2$/Vs) for sample~1 after illumination at
low $T$.
Though both, $n_s$ and $\tau_p$, increase by about 30\%, the
photocurrent remains unchanged, thus confirming the above
arguments. For high temperatures the carrier distribution is
sufficiently well described by the Boltzmann function and hence $S
\propto 1/k_B T$ (see Eq.~(\ref{spin})). Therefore, the current
is proportional with $n_s /k_B T$ and decreases with
increasing temperature   in agreement to experiment (see solid
line in Fig.~\ref{figure4}(a), showing the fit of data to $n_s
/T$, obtained with one ordinate scaling parameter). In the
intermediate range between 25~K and 100~K, such simple analysis
fails. In this temperature range the scattering mechanism changes
from impurity dominated to phonon dominated.
 This transition region is not yet considered theoretically.

The theoretical treatment of the photocurrent contribution due to
the excitation mechanisms (Fig.~\ref{figure1}(b)) was developed
in Ref. 32
and describes the dependencies of
$j_1$ and $j_3$  on magnetic field, polarization, and temperature
quite well.

To describe the polarization independent contribution $j_2$ we
develop the microscopic theory of the magneto-induced photocurrent
caused by energy relaxation. Our treatment is based on the
spin-density-matrix formalism and presented here for acoustic
phonon mediated electron scattering. The energy relaxation of
spin-polarized hot carriers in gyrotropic structures and in the
presence of a magnetic field is accompanied by the generation of
an electric current which is given by
\begin{equation}\label{j_gen}
\bm{j}_{\mathrm{rel}} = 2 e \sum_{\bm{k}\,\bm{k}'}\sum_{s=\pm 1/2}
(\bm{v}_{\bm{k}} -\bm{v}_{\bm{k}'}) \,\tau_p \, f_{\bm{k}'s}
(1-f_{\bm{k}s}) \, w_{\bm{k}s \leftarrow \bm{k}'s} \:,
\end{equation}
where the index $s$  designates the spin state,
$\bm{v}_{\bm{k}}=\hbar\bm{k}/m^*$ is the velocity, $m^*$ the
effective electron mass, $f_{\bm{k}s}$  the distribution function
of carriers in the spin subband $s$, $w_{\bm{k}s \leftarrow
\bm{k}'s}$  the rate of phonon-induced electron scattering, and
the factor 2 in Eq.~(\ref{j_gen}) accounts for the valley
degeneracy in SiGe (001)-grown QW structures. The scattering rate
has the form
%

\begin{eqnarray} w_{\bm{k}s \leftarrow \bm{k}'s}  = 
 \frac{2\pi}{\hbar} \sum_{\bm{q}, \pm} |M_{\bm{k}s,\,\bm{k}'s}|^2 \times\nonumber\\
\times\left(N_{\bm{q}} + \frac12 \pm \frac12 \right)
\delta(\varepsilon_{\bm{k}} - \varepsilon_{\bm{k}'} \pm
\hbar\Omega_{\bm{q}}) \:,
\end{eqnarray}
where $\Omega_{\bm{q}}$ and $\bm{q}$ are the frequency and wave
vector of the phonon involved, $M_{\bm{k}s,\,\bm{k}'s}$ is the
matrix element of electron-phonon interaction, $N_{\bm{q}}$  the
phonon occupation number,
$\varepsilon_{\bm{k}}=\hbar^2\bm{k}^2/2m^*$  the electron kinetic
energy, and the signs ``$\pm$'' correspond to the phonon emission
and absorption.

For the structure symmetry described by the axial C$_{\infty v}$
point group, which is obviously relevant for the SiGe-based QWs
under study, the matrix element of electron-phonon interaction can
be modelled by
\begin{equation}\label{scattering}
M_{\bm{k},\bm{k}'} = \mathcal{A}(q_z) + \mathcal{B}(q_z) [\sigma_x
(k_y+k'_y) - \sigma_y (k_x+k'_x)] \:,
\end{equation}
where $\mathcal{A}(q_z)$ and $\mathcal{B}(q_z)$ are material
parameters determined by the QW structure, $q_z$ is the phonon
wave vector component along the growth direction. Further we
assume that the inequality $q_x, q_y \ll q_z$ is fulfilled for the
typical phonons involved.

We consider heating of the electron gas  by a radiation and that
the electron temperature (same for the both spin subbands)
slightly exceeds the lattice temperature. Then, using the
Boltzmann distribution of the carriers for the non-degenerate case
(the high-temperature range) and in the quasi-elastic
approximation, one derives
\begin{equation}
\label{current2}
j_{\mathrm{rel},x} = 4 e \tau_p S_y \,\xi I \eta /\hbar \:, \;\;
j_{\mathrm{rel},y} = - 4 e \tau_p S_x \,\xi I \eta /\hbar \:,
\end{equation}
where $\xi = \sum_{q_z}A(q_z)B(q_z)\,|q_z|
/\sum_{q_z}A^2(q_z)\,|q_z|$ is a parameter which is determined by
the ratio  of the  spin-dependent and spin-independent parts of
the electron-phonon interaction.

From Eqs.~(\ref{current2}) it follows that, as expected, the
contribution $\bm{j}_{\mathrm{rel}}$ is independent of the
polarization state of radiation and thus it may appear also for unpolarized
radiation.
Thus, for high temperatures, the temperature dependence of
$j_{\mathrm{rel}}$ is described by the simple expression
$j_{\mathrm{rel}} \propto  n_s  / k_B T$ because $\eta \propto
n_s$ and  $S_y \propto 1 / k_B T$. A fit of this function to the
data is shown as solid line in Fig.~\ref{figure4}(b) demonstrating
good agreement. Treatment of the low temperature range needs
allowance for terms of higher order in the in-plane phonon wave
vector and is out of scope of this paper.

In addition to the magnetic field, polarization and temperature
dependencies of the magneto-photocurrent we investigate its
anisotropic properties. For that we vary  the orientation of the
in-plane magnetic field relative to the crystallographic
direction. We observe that the magnitude of the current remains
unchanged within the experimental accuracy  for both $\bm j \,\bot
\,\bm B$ and $\bm j \parallel \bm B$ geometries. This isotropic
behaviour of the current agrees well with the microscopic picture
described above and it can be attributed to the fact that,  in
contrast to zinc-blende structure based QWs, structure inversion
asymmetry is the only possible asymmetry  in conventional SiGe
QWs.

Summarizing all data, we demonstrate that in asymmetric SiGe QWs
spin-dependent scattering  results in a pure spin current and spin
separation. We show that application of an external magnetic field
gives experimental access to investigations of pure spin currents.
The basis of the method is the conversion of a pure spin current
into an electric current by means of a magnetic field induced
equilibrium spin polarization.

\subsection*{Acknowledgements}  This work was supported by the
DFG via Project GA 501/6-2  and Collaborative Research Center
SFB689, the RFBR, programs of the RAS, the HBS, Russian Science
Support Foundation, and in Austria by the ``Fonds zur
F\"{o}rderung der Wissenschaftichen Forschung'', Vienna.


\begin{thebibliography}{99}


\bibitem{Zutic04review} I.~\u{Z}uti\'{c}, J.~Fabian, and S.~Das~Sarma,
Rev. Mod. Phys. \textbf{76}, 323 (2004).

\bibitem{Bychkov84p78} Yu.A.~Bychkov and E.I.~Rashba,
Pis'ma Zh. Eksp. Teor. Fiz. \textbf{39}, 66
(1984) [JETP Lett. \textbf{39}, 78 (1984)].%

\bibitem{Wilamowski2002p195315} Z.~Wilamowski, W.~Jantsch, H.~Malissa,  and U.~R\"{o}ssler,
Phys. Rev. B \textbf{66}, 195315 (2002).

\bibitem{Tyryshkin2003}A.M.~Tyryshkin, S.A.~Lyon, W.~Jantsch, and F.~Sch\"{a}ffler,
Phys. Rev. Lett. \textbf{94}, 126802 (2005).

\bibitem{Ganichev2002b} S.D.~Ganichev, U.~R\"{o}ssler, W.~Prettl, E.L.~Ivchenko,
V.V.~Bel'kov, R.~Neumann, K.~Brunner,  and G.~Abstreiter, Phys.
Rev. B \textbf{66}, 075328 (2002).

\bibitem{Belkov2003} V.V.~Bel'kov, S.D.~Ganichev, P.~Schneider,
D.~Schowalter, U.~R\"{o}ssler, W.~Prettl, E.L.~Ivchenko,
R.~Neumann, K.~Brunner,  and G.~Abstreiter,
J.~Supercond.: Incorp. Novel Magn. \textbf{16}, 415 (2003).

\bibitem{Tahan2002} C.~Tahan, M.~Friesen,  and R.~Joynt,
Phys. Rev. B   \textbf{66}, 035314 (2002).

\bibitem{Fanciulli2003} M.~Fanciulli, P.~H\"{o}fer,  and A.~Ponti,
Physica B \textbf{340\,-342}, 895 (2003).

\bibitem{Wilamowski2004p35328} Z.~Wilamowski and  W.~Jantsch,
Phys. Rev. B \textbf{69}, 035328 (2004).

\bibitem{Glazov2004} M.M.~Glazov,
Phys. Rev. B   \textbf{70}, 195314 (2004).

\bibitem{Tahan2005} C.~Tahan and  R.~Joynt,
Phys. Rev. B   \textbf{71}, 075315 (2005).

\bibitem{Sherman2003}   E.Ya.~Sherman,
Appl. Phys. Lett.   \textbf{82}, 209 (2003).

\bibitem{Jantsch2002p504}    W.~Jantsch,  Z.~Wilamowski,
N.~Sandersfeld, M.~Muhlberger,  and F.~Sch\"{a}ffler,
Physica E \textbf{13}, 504 (2002).

\bibitem{Kiselev2003}    F.A.~Baron, A.A.~Kiselev, H.D.~Robinson, K.W.~Kim, K.L.~Wang,
 and E.~Yablonovitch,
Phys. Rev. B   \textbf{68}, 195306 (2003).

\bibitem{Malissa2004p1739}
H.~Malissa, W.~Jantsch, M.~M\"{u}hlberger, F.~Sch\"{a}ffler,
Z.~Wilamowski, M.~Draxler,  and P.~Bauer,
Appl. Phys. Lett. \textbf{85}, 1739 (2004).

\bibitem{Truitt2004}  J.L.~Truitt \textit{et al}.,
cond-mat/0411735 (2004).

\bibitem{Wilamowski2006} Z.~Wilamowski, H.~Malissa, F.~Sch\"{a}ffler, and
W.~Jantsch,
cond-mat/0610046 (2006).

\bibitem{PRL01}S.D.~Ganichev, E.L.~Ivchenko, S.N.~Danilov, J.~Eroms,
W.~Wegscheider, D.~Weiss,  and W.~Prettl,
Phys. Rev. Lett. \textbf{86}, 4358 (2001).

\bibitem{Nature02} S.D.~Ganichev, E.L.~Ivchenko,
V.V.~Bel'kov, S.A.~Tarasenko, M.~Sollinger, D.~Weiss,
W.~Wegscheider,  and W.~Prettl,
Nature (London) \textbf{417}, 153 (2002).

\bibitem{Ganichev03p935} S.D.~Ganichev and W.~Prettl,
J. Phys.: Condens. Matter {\bf 15}, R935 (2003).

\bibitem{Belkov283p2003} V.V.~Bel'kov, S.D.~Ganichev, Petra~Schneider, C.~Back,
M.~Oestreich, J.~Rudolph, D.~H{\"a}gele, L.E.~Golub,
W.~Wegscheider,  and W.~Prettl,
Solid State Commun. \textbf{128}, 283 (2003).

\bibitem{Ivchenkobook2} E.L.~Ivchenko, \textit{ Optical
Spectroscopy of Semiconductor Nanostructures} (Alpha Science Int.,
Harrow,  2005).

\bibitem{Bieler05}   M.~Bieler, N.~Laman, H.M.~van~Driel,  and A.L.~Smirl,
Appl. Phys. Lett. \textbf{86}, 061102 (2005).

\bibitem{GanichevPrettl} S.D.~Ganichev and  W.~Prettl, {\it Intense Terahertz Excitation of Semiconductors}
(Oxford University Press, Oxford, 2006).

\bibitem{Yang06} C.L.~Yang, H.T.~He, Lu~Ding, L.J.~Cui, Y.P.~Zeng, J.N.~Wang,  and W.K.~Ge,
Phys. Rev. Lett. \textbf{96}, 186605 (2006).

\bibitem{Huebner2003}  J.~H{\"u}bner, W.W.~R{\"u}hle, M.~Klude, D.~Hommel,
R.D.R.~Bhat, J.E.~Sipe,  and H.M.~van~Driel,
Phys. Rev. Lett. \textbf{90}, 216601 (2003).

\bibitem{Stevens2003} M.J.~Stevens, A.L.~Smirl, R.D.R.~Bhat, A.~Najimaie,
J.E.~Sipe,  and H.M.~van~Driel,
Phys. Rev. Lett. \textbf{90}, 136603 (2003).

\bibitem{Sipe} H.~Zhao, X.~Pan, A.L.~Smirl, R.D.R.~Bhat, A.~Najmaie, J.E.~Sipe,
and H.M.~van~Driel,
Phys. Rev. B \textbf{72}, 201302 (2005).

\bibitem{TI_jetplett} S.A.~Tarasenko and  E.L.~Ivchenko, Pis'ma Zh.
Eksp. Teor. Fiz. \textbf{81}, 292 (2005) [JETP Lett. \textbf{81},
231 (2005)].

\bibitem{Awschalom} Y.~Kato, R.C.~Myers, A.C.~Gossard,  and  D.~Awschalom,
Science \textbf{306}, 1910 (2004).

\bibitem{Wunderlich} J.~Wunderlich, B.~Kaestner, J.~Sinova,  and T.~Jungwirth,
Phys. Rev. Lett. \textbf{94}, 047204 (2005).

\bibitem{Ganichev06zerobias} S.D.~Ganichev  \textit{et al}.,
Nature Phys. \textbf{2}, 609 (2006).

\bibitem{bulli} E.L.~Ivchenko and  G.E.~Pikus, Izv. Akad. Nauk
SSSR  (ser. fiz.) \textbf{47}, 2369 (1983) [Bull. Acad. Sci. USSR,
Phys. Ser. \textbf{47}, 81 (1983)].

\bibitem{ac_field} S.A.~Tarasenko,
Phys. Rev. B \textbf{73}, 115317 (2006).

\bibitem{Dyakonov71} M.I.~D'yakonov and  V.I.~Perel', Pis'ma Zh. Eksp. Teor. Fiz. \textbf{13}, 657 (1971)
[JETP Lett. \textbf{13}, 467 (1971)].

\bibitem{Hirsch99}  J.E.~Hirsch, Phys. Rev. Lett. \textbf{83}, 1834 (1999).

\bibitem{Tarasenko} S.A.~Tarasenko,
Pis'ma Zh. Eksp. Teor. Fiz. \textbf{84}, 233 (2006) [JETP Lett. \textbf{84}, 199 (2006)].

\bibitem{footnote1} We consider here only the spin-dependent
contribution induced by heteropotential asymmetry. Other terms are
negligible in SiGe structures.

\bibitem{footnote}   This is valid (i) for both the  relaxation (Fig.~\ref{figure1}(a))
and the excitation (Fig.~\ref{figure1}(b)) mechanisms in the case
of a non-degenerate distribution of carriers and (ii) for the
excitation mechanism in the case of a degenerate distribution with
the Fermi energy $\varepsilon_F$ provided $\hbar\omega <
\varepsilon_F$ or $\hbar\omega \gg \varepsilon_F$.

\bibitem{purespincurrent}
Strictly speaking, the pure spin current, i.e. the flux of
electron spins, is described by a second-rank pseudotensor with
the components $J^{\alpha}_{\beta}$ giving the flow in the
$\beta$-direction of spins oriented along $\alpha$, with $\alpha$
and $\beta$ being the Cartesian coordinates. Then, the electric
current induced by imbalance of the pure spin photocurrent in
magnetic field is given by
$j_{\beta}=4e\sum_{\alpha}S_{\alpha}J^{\alpha}_{\beta}$.

\bibitem{MPGE_jpcm} V.V.~Bel'kov \textit{et al}.,
J. Phys.: Condens. Matter {\bf 17}, 3405 (2005).

\bibitem{Seeger} K.~Seeger, \textit{Semiconductor Physics} (Springer,
Wien, 1997).


\end{thebibliography}
\end{document}